\documentclass[a4paper,11pt]{article}
\usepackage{jcappub}

\usepackage{subfigure}
\usepackage[subfigure]{graphfig}
\usepackage{color}
\usepackage[toc,page]{appendix}

\usepackage{graphicx}
\usepackage{comment}
\usepackage{array,amsmath,amsthm,feynmf}
\usepackage{bm}
\usepackage{slashed}
\usepackage{times}
\usepackage{epsfig}
\usepackage{cancel}
\usepackage{amssymb}
\usepackage{textcomp}
\usepackage{mathrsfs}
\usepackage{physics}

\usepackage[section]{placeins}
\makeatletter
\AtBeginDocument{%
  \expandafter\renewcommand\expandafter\subsection\expandafter{%
    \expandafter\@fb@secFB\subsection
  }%
}
\makeatother

\title{\boldmath Inflation in the Mixed Higgs-$R^2$ Model }

\author[1,2]{Minxi He,}
\author[2,3]{Alexei A. Starobinsky,}
\author[1,2,4]{Jun'ichi Yokoyama}

\affiliation[1]{Department of Physics, Graduate School of Science,\\ The University of Tokyo, Tokyo 113-0033, Japan}
\affiliation[2]{Research Center for the Early Universe (RESCEU), Graduate School
	of Science,\\ The University of Tokyo, Tokyo 113-0033, Japan}
\affiliation[3]{L. D. Landau Institute for Theoretical Physics, Moscow 119334, Russia}
\affiliation[4]{Kavli Institute for the Physics and Mathematics
	of the Universe (Kavli IPMU), WPI, UTIAS,\\
	The University of Tokyo, Kashiwa, Chiba 277-8568, Japan}

\emailAdd{hemxzero@resceu.s.u-tokyo.ac.jp}
\emailAdd{alstar@landau.ac.ru}
\emailAdd{yokoyama@resceu.s.u-tokyo.ac.jp}

\abstract{We analyze a two-field inflationary model consisting of the Ricci scalar squared ($R^2$) term and the standard Higgs field non-minimally coupled to gravity in addition to the Einstein $R$ term.  Detailed analysis of the power spectrum of this model with mass hierarchy is presented, and we find that one can describe this model as an effective single-field model in the slow-roll regime with a modified sound speed. The scalar spectral index predicted by this model coincides with those given by the $R^2$ inflation and the Higgs inflation implying that there is a close relation between this model and the $R^2$ inflation already in the original (Jordan) frame.  For a typical value of the self-coupling of the standard Higgs field at the high energy scale of inflation, the role of the Higgs field in parameter space involved is to modify the scalaron mass, so that the original mass parameter in the $R^2$ inflation can deviate from its standard value when non-minimal coupling between the Ricci scalar and the Higgs field is large enough. }

\begin{document}
\maketitle
\flushbottom

\section{Introduction} \label{sec:intro}

A number of single-field models have been proposed \cite{Starobinsky:1980te,Sato:1980yn,Guth:1980zm,Linde:1981mu,Albrecht:1982wi,Sato:2015dga} since 1980s, some of them are in good agreement with the observation of cosmic microwave background (CMB) \cite{Ade:2015lrj}, such as the  $R+R^2$ inflationary model (the $R^2$ one for brevity)  \cite{Starobinsky:1980te} which is often called the Starobinsky model, and the original Higgs inflationary model\cite{CervantesCota:1995tz,Bezrukov:2007ep,Barvinsky:2008ia}
in which the scalar field is strongly non-minimally coupled to the Ricci scalar \footnote{See 
\cite{Kamada:2012se} for a summary of all variants of the Higgs inflationary model}.   The $R^2$ added to the Einstein-Hilbert action yields an effective dynamical scalar field, scalaron realizing a quasi-de Sitter stage in the early universe while the Higgs boson in the standard model, with the help of non-minimal coupling to gravity, $ \xi \chi^2 R $, plays an essential role as an inflaton to drive inflation in the Higgs inflationary model.  Both models produce the same spectral spectral index of primordial scalar (adiabatic density) perturbations which is supported by recent CMB observations.  Meanwhile, the tensor-to-scalar ratio given by these two models has an amplitude though small, but still hopefully detectable in the future. 

Due to the excellent performance of the $R^2$ inflation and the Higgs inflationary model, it is natural and more realistic to consider the extension of such single-field models to multi-field inflation by the combination of them which we consider in this paper.  Multi-field inflation is a class of cosmological inflationary models with a de Sitter stage produced by more than one effective scalar fields among which two-field models constitute a special case. In multi-field inflationary models, only one linear combination of the scalar fields is responsible for the inflationary stage and consequently quantum fluctuations produced in this direction serve as adiabatic perturbations which finally grow to become the seeds of inhomogeneities seen in CMB temperature anisotropy and polarization and producing the large scale structure and compact objects in the universe. The other independent combinations are, on the other hand, responsible for production of isocurvature perturbations \cite{Kodama:1985bj} and some other possible features \cite{Chen:2009we}. Isocurvature modes represent the unique feature of multi-field models distinguishing them from single-field ones. They can survive to the present only under special conditions \cite{Polarski:1994rz}. Also, in the presence of non-minimal coupling, recent research \cite{Ema:2016dny} points out that the preheating process after inflation becomes much more violent than the case without it. 

In this paper, we investigate Higgs-$ R^2 $ inflation, namely the combination of the Higgs inflation and the $R^2$ inflation, in a certain part of the parameter space.  For realistic values of the Higgs self coupling we find the presence of mass hierarchy and the appearance of effectively single-field slow-roll inflation in the original Jordan frame. We write down the effective single-field action to quadratic level for this model and use it to calculate the power spectrum of curvature perturbations.  We find that this two-field model can be treated as an effective $R^2$ inflation with a modified scalaron mass. In Sec~\ref{sec:L-EOM}, we introduce the basic details of the model. We calculate the power spectrum in Sec~\ref{sec:srcp} and discuss the effective $R^2$ inflation in Sec~\ref{sec:equiv}. Our conclusions and outlook are presented in Sec~\ref{sec:concloutl}. 

\section{Lagrangian and Equations of Motion}\label{sec:L-EOM}

The action considered here is given in the original Jordan frame, where the space-time metric is denoted as $\hat{g}_{\mu\nu}$, by 
\begin{align}\label{action:j}
	 S_{\text{J}} &=\int d^4 x \sqrt{-\hat{g}} \left[\frac{M_p^2}{2} \hat{R}+\frac{1}{2}\xi \chi^2 \hat{R} +\frac{M_p^2}{12 M^2} \hat{R}^2 -\frac{1}{2}\hat{g}^{\mu \nu}\hat{\nabla}_{\mu}\chi \hat{\nabla}_{\nu}\chi -\frac{\lambda}{4} \chi^4 \right] \\
	 &=\int d^4 x \sqrt{-\hat{g}} \left[F(\chi, \hat{R})-\frac{1}{2}\hat{g}^{\mu \nu}\hat{\nabla}_{\mu}\chi \hat{\nabla}_{\nu}\chi \right]
\end{align} 
where $M_p \equiv(8\pi G)^{-1/2}$ and $ \chi $ is a singlet scalar field, a simplified model of the Standard Model Higgs boson. We neglect its interaction to gauge fields. Here $ F(\chi,\hat{R}) $ is defined by 
\begin{align}
   F(\chi, \hat{R}) \equiv \frac{M_p^2}{2} \hat{R}+\frac{1}{2}\xi \chi^2 \hat{R} +\frac{M_p^2}{12 M^2} \hat{R}^2 -\frac{\lambda}{4} \chi^4 ~.
\end{align}
This action was recently considered in \cite{Ema:2017rqn,Wang:2017fuy}. $\chi$ has a non-minimal coupling term with the Ricci scalar. We take the sign of the non-minimal coupling constant $\xi$ such that the conformal coupling corresponds to $ \xi=-1/6 $. 

Defining the scalaron field as \cite{Maeda:1988ab,Maeda:1987xf} 
\begin{align}
   \label{def:psi}\sqrt{\frac{2}{3}}\frac{\psi}{M_p}\equiv \ln (\frac{2}{M^2_p}\abs{\frac{\partial F}{\partial \hat{R}}})~,
\end{align}
and performing a conformal transformation
\begin{align}
   g_{\mu \nu} (x)=e^{\sqrt{\frac{2}{3}}\frac{\psi(x)}{M_p}} \hat{g}_{\mu \nu}(x)~,
\end{align}
we can transform the original action \eqref{action:j} into the one in the
Einstein frame and express the new action in terms of the new scalar fields as
\begin{align}
   S_{\text{E}} =\int d^4 x \sqrt{-g} \left[\frac{M_p^2}{2} R -\frac{1}{2}g^{\mu \nu}\nabla_{\mu}\psi \nabla_{\nu}\psi -\frac{1}{2} e^{-\sqrt{\frac{2}{3}}\frac{\psi}{M_p}} g^{\mu \nu}\nabla_{\mu}\chi \nabla_{\nu}\chi -U(\psi, \chi) \right] ~,
\end{align}
where the potential is expressed as 
\begin{align}\label{potential}
   U(\psi, \chi) \equiv \frac{\lambda}{4} \chi^4 e^{-2\sqrt{\frac{2}{3}}\frac{\psi}{M_p}} + \frac{3}{4} M_p^2 M^2 e^{-2\sqrt{\frac{2}{3}}\frac{\psi}{M_p}} \left(e^{\sqrt{\frac{2}{3}}\frac{\psi}{M_p}} -1 -\frac{1}{M_p^2} \xi \chi^2\right)^2 ~.
\end{align}
In addition to the metric and Higgs field $\chi$, $\psi$ is the third dynamical field which originates from the $R^2$ term. Note that $\psi$ shows up only in the exponent with an $\mathcal{O}(1)$ numerical factor, so that large values of $\psi$ will significantly suppress the terms with higher orders in $\exp (-\sqrt{2/3}\psi/M_p)$. 

The kinetic terms of the two scalar fields in the Einstein frame are coupled. This means that the field space spanned by these two fields is not flat. Following \cite{Sasaki:1995aw,Achucarro:2010da}, we introduce an induced metric of the field space and rewrite this system in a more compact way as 
\begin{align}
   S_{\text{E}}=\int d^4 x \sqrt{-g} \left[\frac{M_p^2}{2} R -\frac{1}{2}h_{ab}g^{\mu \nu}\nabla_{\mu}\phi^a \nabla_{\nu}\phi^b -U(\phi) \right],
\end{align}
where 
\begin{align}
	\phi^1=\psi~,\quad \phi^2=\chi~, \quad 
	h_{ab}=h_{ab}(\psi)=
	\begin{pmatrix}
	1 & 0 \\ 
	0 & e^{-\sqrt{\frac{2}{3}}\frac{\psi}{M_p}}. \\
	\end{pmatrix}
\end{align}
Here the Latin indices $ a,b=1,2 $ represent components in field space and the Greek indices $ \mu,\nu=0,1,2,3 $ denote space-time components. 

We take the spatially flat Robertson-Walker metric $ ds^2=-dt^2+a^2(t)\delta_{ij}dx^idx^j $ 
as the background {$i,j=1,2,3$) and split all fields into homogeneous background parts and small space-time-dependent perturbations, 
\begin{align}
	\phi^a(\mathbf{x},t)=\phi^a_0(t)+\delta\phi^a(\mathbf{x},t) ~,
\end{align}
incorporating scalar metric perturbations in the spatially flat gauge.
Then equations of motion for both background and perturbations are given as follows.  
\begin{align}
   &\label{fried}H^2=\frac{1}{3M^2_p}\left[ \frac{1}{2}h_{ab}\dot{\phi}^a\dot{\phi}^b+U(\phi) \right]~, \\
   &\label{eom:bg} \frac{D\dot{\phi}^a_0}{dt}+3H\dot{\phi}^a_0+h^{ab}U_{,b}=0~, \\
   \label{eom:pert} \frac{D^2\delta\phi^a_{\mathbf{k}}}{dt^2}+3H\frac{D\delta\phi^a_{\mathbf{k}}}{dt}-R^a_{\; bcd}\dot{\phi}^b_0&\dot{\phi}^c_0\delta\phi^d_{\mathbf{k}}+\frac{k^2}{a^2}\delta\phi^a_{
   \mathbf{k}}+U^{;a}_{\;\;;b}\delta\phi^b_{\mathbf{k}} =\frac{1}{a^3}\frac{D}{dt}\left(\frac{a^3}{H}\dot{\phi}^a_0\dot{\phi}^b_0\right)h_{bc}\delta\phi^c_{\mathbf{k}} ~,
\end{align}
where $ D X^a= d X^a+\Gamma^a_{bc}X^bd\phi^c_0 $ is analogous to the directional derivative in curved spacetime and $ \Gamma^a_{bc}=\frac{1}{2}h^{ad}(\partial_bh_{cd}+\partial_ch_{db}-\partial_dh_{bc}) $ is the Christoffel symbol for the curved field space. It is easy to show that $ \frac{D}{dt}=\dot{\phi}^a_0\nabla_{a} $. Note that the equations of motion for perturbations have already been transformed into those for spatial Fourier modes. The equations of motion of the scalar fields can be regarded as modified geodesic equations in the curved field space. The first term in \eqref{eom:bg} is just the ordinary geodesic equation while the second and the third term represent modifications from cosmic expansion and the scalar field potential, respectively. 
Correspondingly, the field perturbation equations \eqref{eom:pert} can be regarded as geodesic deviation. Their equations of motion are also modified by cosmic expansion and the potential. 
With all the effects taken into account, the trajectory of the fields traces neither the geodesics in the curved field space nor the bottom of the valley of the potential as postulated in \cite{Wang:2017fuy}. Note that since generally the trajectory take turns during inflation, we also expect that there will be effects due to the turning. 

\section{Slow-Roll Inflation and Curvature Perturbations}\label{sec:srcp}

In this paper, we mainly focus on the parameter regime where $ \xi>0 $ and fix the self coupling
at a typical value $ \lambda=0.01 $ from phenomenology. 
We will also briefly discuss the situation when $ \xi<0 $ at the end. 

\subsection{Features of the Potential}

Here we give two examples for different combinations of $ \xi $ and $ M $ in Figure~\ref{fig:pot}. The potential \eqref{potential} is invariant under $ \chi\rightarrow -\chi $. One can calculate the effective mass of the Higgs field, $ m^2_{\chi} $, by taking derivatives of the potential. The dominant contribution in small $ \chi $ regime comes from a term proportional to $ \xi $, $ -3\xi M^2 \exp (-\sqrt{2/3}\psi/M_p) $. For positive $ \xi $, Higgs field obtains a negative $ m^2_{\chi} $ around the origin where its amplitude will grow exponentially. In large $ \chi $ regime, $ m^2_{\chi} $ is dominated by a term proportional to $ \chi^2 $, $ 3\lambda(1+3\xi^2M^2/\lambda M^2_p)\exp(-2\sqrt{2/3}\psi/M_p)\chi^2 $, whose coefficient is always positive. These properties imply the existence of a local minimum on the potential for a given $ \psi $ which corresponds to the valleys in Figure~\ref{fig:pot}. Thus, independent of the initial position of $ \chi $, with a large $ \xi $, the Higgs field will quickly fall into one of the valleys and evolve around the local minimum. If $ \xi $ takes a small value, i.e. $ m^2_{\chi} $ is small, $ \chi $ direction will become flatter. In this case, if the initial conditions start from a large $ \chi $ value, it is possible for Higgs field to slowly roll down the potential wall which is similar to the
situation discussed in \cite{Pi:2017gih}. As for $ \psi $ direction, it is always flat in large $ \psi $ regime so that it has similar behavior to the scalaron in the $R^2$ inflation. As we shall see later, there is a turning in the trajectory which can affect the sound speed of the curvature perturbations during inflationary phase. The angular velocity at this turning is not large in the parameter regime we consider here, though. After the end of inflation, the fields will oscillate around the global minimum of the potential at $ (\chi,\psi)=(0,0) $ where reheating is expected to happen. According to the recent work \cite{Ema:2016dny}, the particle production during preheating will be violent due to the appearance of non-minimal coupling between Higgs and gravity. 

\begin{figure}[h]
	\centering
	\begin{subfigure}
		\centering
		\label{fig:pot-1000}
		\includegraphics[width=.48\textwidth]{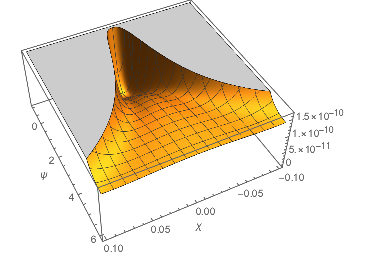}
	\end{subfigure}
	\begin{subfigure}
		\centering
		\label{fig:pot-3000}
		\includegraphics[width=.48\textwidth]{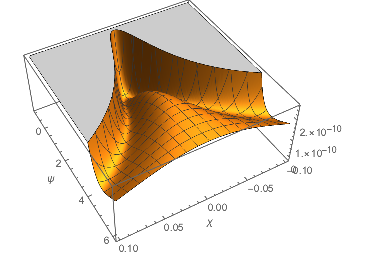}
	\end{subfigure}
	\caption{Left: $ \xi=1000 $, $ M/M_p=1.4\times 10^{-5} $. Right: $ \xi=3000 $, $ M/M_p=1.9\times 10^{-5}$. The field values have been normalized by $ M_p $. The value of $ \xi $ mainly controls the position of the valleys while $ M $ determines the height of the plateau in the middle part and the depth of the valley.}
	\label{fig:pot}
\end{figure}

\subsection{Slow-Roll Inflation}

As mentioned in previous sections, the evolution trajectory of two scalar fields are affected by the curved nature of the field space, the potential shape and the expansion of the universe. Thus, it would be more convenient to discuss the features of this trajectory by defining unit vectors $ T^a $ and $ N^a $ \cite{Achucarro:2012yr} as
\begin{align}
	T^a\equiv\frac{\dot{\phi}^a_0}{\dot{\phi}_0}&=\frac{1}{\sqrt{\dot{\psi}^2+e^{-\sqrt{\frac{2}{3}}\frac{\psi}{M_p}}\dot{\chi}^2}}(\dot{\psi},\dot{\chi}),\\
	\dot{\theta}N^a&\equiv -\frac{DT^a}{dt}~,
\end{align}
which are tangent and normal to the trajectory, respectively. Here we denote $ \dot{\phi}^2_0\equiv h_{ab}\dot{\phi}^a_0\dot{\phi}^b_0 $ and $ \dot{\theta} $ is the angular velocity describing the turning in the trajectory which,  according to the normalization condition, is given by 
\begin{align}
	\dot{\theta}^2&=h_{ab}\frac{DT^a}{dt}\frac{DT^b}{dt}\\
	&=e^{\sqrt{\frac{2}{3}}\frac{\psi}{M_p}}\frac{\left(\frac{\partial U}{\partial\chi}\dot{\psi}-e^{-\sqrt{\frac{2}{3}}\frac{\psi}{M_p}}\frac{\partial U}{\partial\psi}\dot{\chi}\right)^2}{\left(\dot{\psi}^2+e^{-\sqrt{\frac{2}{3}}\frac{\psi}{M_p}}\dot{\chi}^2\right)^2}~,
\end{align}
so that $ N^a $ is explicitly given by 
\begin{align}
	N^a=\frac{e^{\sqrt{\frac{1}{6}}\frac{\psi}{M_p}}}{\left(\dot{\psi}^2+e^{-\sqrt{\frac{2}{3}}\frac{\psi}{M_p}}\dot{\chi}^2\right)^{1/2}}(-e^{-\sqrt{\frac{2}{3}}\frac{\psi}{M_p}}\dot{\chi},\dot{\psi})~.
\end{align}
We define the slow-roll parameters analogous to the single-field case as 
\begin{align}
	\epsilon&\equiv -\frac{\dot{H}}{H^2}=\frac{\dot{\phi}^2_0}{2M^2_pH^2}~, \\
	\eta^a &\equiv -\frac{1}{H\dot{\phi}_0}\frac{D\dot{\phi}^a_0}{dt}~.
\end{align}
Note that $ \eta^a $ is no longer a scalar but a vector which means that one needs two different $ \eta $s to describe the evolution of these two different directions. Using the unit vectors, one can easily obtain an $ \eta $ for each direction as 
\begin{align}
	\eta^a&=\eta_{||}T^a+\eta_{\perp}N^a~, \\
	\eta_{||}&\equiv -\frac{\ddot{\phi}_0}{H\dot{\phi}_0}~, \\
	\eta_{\perp}&\equiv \frac{U_N}{\dot{\phi}_0H}~,
\end{align}
where $ U_N\equiv N^aU_{,a} $. Slow-roll inflation requires that $ \epsilon\ll 1 $ and $ \eta_{||}\ll 1 $. Note that the slow-roll requirement does not impose any constraint on  $ \eta_{\perp} $ which means that it can be large. Then the angular velocity $ \dot{\theta} $ can also be expressed in terms of the slow-roll parameter as 
\begin{align}
	\dot{\theta}=H \eta_{\perp}~.
\end{align}
Since $ \eta_{\perp} $ can be large, we may expect $ \dot{\theta} $ to be large as well. However, this does not spoil the validity of the effective field theory used below \cite{Achucarro:2012yr} as long as the adiabatic condition, $ |\ddot{\theta}/\dot{\theta}^2|\ll M_{\text{eff}} $, is satisfied. 

We now consider perturbations in this formalism. In flat gauge, the comoving curvature perturbation and the isocurvature perturbation are defined as \cite{Achucarro:2012yr} 
\begin{align}
	\mathcal{R}&\equiv -\frac{H}{\dot{\phi}_0}\delta\phi^aT_a~,\\
	\mathcal{F}&\equiv N_a\delta\phi^a~.
\end{align}
Expanding the perturbed action to second order, we find 
\begin{align}
	S_2=\frac{1}{2}\int d^4x~ a^3\left[\frac{\dot{\phi}^2_0}{H^2}\dot{\mathcal{R}}^2-\frac{\dot{\phi}^2_0}{H^2}\frac{(\nabla\mathcal{R})^2}{a^2}+\dot{\mathcal{F}}^2-\frac{(\nabla\mathcal{F})^2}{a^2}-M^2_{\text{eff}}\mathcal{F}^2-4\dot{\theta}\frac{\dot{\phi}_0}{H}\dot{\mathcal{R}}\mathcal{F}\right]
\end{align}
from which it is clear that the curvature perturbations evolve along the light (massless) direction while the isocurvature modes have an effective mass $ M^2_{\text{eff}}=U_{\text{NN}}+M^2_p\epsilon H^2 R_{ab}h^{ab} -\dot{\theta}^2 $ where $ U_{NN}\equiv N^aN^b\nabla_a\nabla_b U $. The explicit form of $ U_{NN} $ and $ M^2_{\text{eff}} $ is given by 
\begin{align}
 \begin{split}
	U_{NN}=\frac{1}{\dot{\psi}^2+e^{-\sqrt{\frac{2}{3}}\frac{\psi}{M_p}}\dot{\chi}^2}&\left(e^{-\sqrt{\frac{2}{3}}\frac{\psi}{M_p}}\dot{\chi}^2\frac{\partial^2U}{\partial\psi^2}+e^{\sqrt{\frac{2}{3}}\frac{\psi}{M_p}}\dot{\psi}^2\frac{\partial^2U}{\partial\chi^2}\right.\\ 
	&\left.-2\dot{\psi}\dot{\chi}\frac{\partial^2U}{\partial\psi\partial\chi}-\frac{1}{\sqrt{6}}\dot{\psi}^2\frac{\partial U}{\partial\psi}-\sqrt{\frac{2}{3}}\dot{\psi}\dot{\chi}\frac{\partial U}{\partial\chi} \right)~,
 \end{split}\\
 \begin{split}
 M^2_{\text{eff}}=\frac{1}{\dot{\psi}^2+e^{-\sqrt{\frac{2}{3}}\frac{\psi}{M_p}}\dot{\chi}^2}&\left(e^{-\sqrt{\frac{2}{3}}\frac{\psi}{M_p}}\dot{\chi}^2\frac{\partial^2U}{\partial\psi^2}+e^{\sqrt{\frac{2}{3}}\frac{\psi}{M_p}}\dot{\psi}^2\frac{\partial^2U}{\partial\chi^2}-2\dot{\psi}\dot{\chi}\frac{\partial^2U}{\partial\psi\partial\chi}-\frac{1}{\sqrt{6}}\dot{\psi}^2\frac{\partial U}{\partial\psi}\right.\\
 \left.-\sqrt{\frac{2}{3}}\dot{\psi}\dot{\chi}\frac{\partial U}{\partial\chi} \right)&-\frac{\dot{\psi}^2 +e^{-\sqrt{\frac{2}{3}}\frac{\psi}{M_p}}\dot{\chi}^2}{6M^2_p}-e^{\sqrt{\frac{2}{3}}\frac{\psi}{M_p}}\frac{\left(\frac{\partial U}{\partial\chi}\dot{\psi}-e^{-\sqrt{\frac{2}{3}}\frac{\psi}{M_p}}\frac{\partial U}{\partial\psi}\dot{\chi}\right)^2}{\left(\dot{\psi}^2+e^{-\sqrt{\frac{2}{3}}\frac{\psi}{M_p}}\dot{\chi}^2\right)^2}
 ~,
 \end{split}
\end{align}
respectively. Thus, light modes and massive modes are separated. Integrating out the high energy degrees of freedom as in \cite{Achucarro:2012yr}, the massive modes, $ \mathcal{F} $, are completely determined by the massless modes, $ \mathcal{R} $, so that one gets its effective action to quadratic order, 
\begin{align}
	\mathcal{F}&=-\frac{\dot{\phi}_0}{H}\frac{2\dot{\theta}\dot{\mathcal{R}}}{k^2/a^2+M^2_{\text{eff}}}~, \\
	S_{\text{eff}}=\frac{1}{2}\int &d^4x~a^3\frac{\dot{\phi}^2_0}{H^2}\left[\frac{\dot{\mathcal{R}}^2}{c^2_s(k)}-\frac{k^2\mathcal{R}^2}{a^2}\right]~.
\end{align}
The appearance of the turning gives corrections to the sound speed, $ c^{-2}_s (k)=1+4\dot{\theta}^2/(k^2/a^2+M^2_{\text{eff}})$ which is exact unity in single-field models. Therefore, the effective action obtained for curvature perturbations is that of a single-field theory with a modified sound speed. When $ \dot{\theta}^2 $ is close to $ U_{NN} $, $ c_s^{-2}\gg 1 $, the effect of the turn in the trajectory becomes significant, so that the sound speed is largely modified. However, in the region of the parameter space we consider, the modification is not significant. 

In this action, only the adiabatic mode appears but it does not mean that the heavy mode has no influence on the evolution of the adiabatic mode. Both light and heavy modes have high energy and low energy contributions. Integrating out the high energy part to get the low energy effective theory does not mean decoupling between light and heavy modes. As long as a turning exists, adiabatic and isocurvature modes couple with each other 
and the isocurvature mode is forced to oscillate coherently with the light field at low frequency \cite{Achucarro:2012yr}. In the slow-roll regime, $ \dot{\theta} $ is automatically small and slowly changing in time, so that approximately we can quantize the quadratic action considering the sound speed as a constant close to unity. As a result, the power spectrum is just 
\begin{align}
	&\Delta^2_{\mathcal{R}}=\frac{k^3}{2\pi^2}\mathcal{P}_{\mathcal{R}}(k)~,\\
	\mathcal{P}_{\mathcal{R}}(k)=\frac{H^2}{4c_s k^3}(1&+c^2_sk^2\tau^2)\frac{2H^2}{\dot{\phi}^2_0}\xrightarrow[\text{large scale}]{}\frac{H^2}{4c_sk^3}\frac{1}{\epsilon}~,
\end{align}
which gives the scalar index and the scalar-to-tensor ratio 
\begin{align}
	n_{\text{s}}-1\equiv &\frac{d\ln \Delta^2_{\mathcal{R}}}{d\ln k}\approx 2\eta_{||}-4\epsilon\\
	&r\approx 16\epsilon c_s~.
\end{align}
These results are just like those in single-field models with modification from the non-trivial sound speed. However, as mentioned already, large modification is not expected because in the slow-roll regime as well as the region of the parameter space we consider, the sound speed does not deviate from unity too much. 

\subsection{Predictions for observations}

As we can see above, the dynamics as well as power spectrum are determined by three parameters, Higgs self-coupling $ \lambda $, non-minimal coupling $ \xi $ and scalaron mass $ M $. 
\begin{figure}[h]
	\centering
	\includegraphics[width=.7\textwidth]{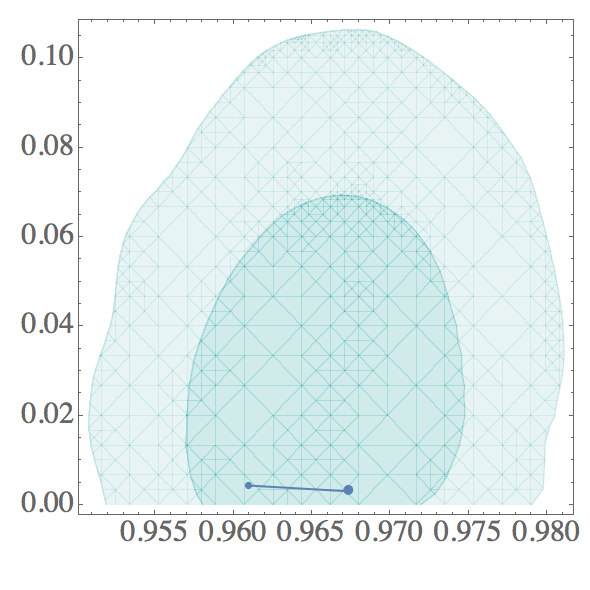}
	\caption{$ \lambda=0.01, \mathcal{P}_{\mathcal{R}}(k)\simeq2\times10^{-9} $~. Results given by different combinations of $ \xi $ and $ M $ are completely degenerate with each other and coincide with predictions given by the $R^2$ inflation. The left small dot is for 50 e-foldings and the right big one is 60 e-foldings. }
	\label{fig:n_stor}
\end{figure}
Fixing $ \lambda=0.01 $ and the amplitude of curvature perturbation at the pivot scale to be $ 2\times 10^{-9} $, we choose several groups of $ \xi $ and $ M $ to calculate $ n_{\text{s}} $ and $ r $. All the results are completely degenerate (Shown in Figure~\ref{fig:n_stor}) with those of the $R^2$ inflation. 
This should not be surprising because this two-field model is built from the $R^2$ inflation and the Higgs inflation both of which give predictions staying right at the center of the famous $n_s-r$ plot from Planck's observational data in 2015 \cite{Ade:2015lrj}. Also, due to the presence of mass hierarchy and considering slow-roll inflation, the effective theory of this model reduces to a single-field model with slightly modified sound speed as one can see above. Therefore, we should not expect this model to give predictions which largely deviate from those of the Higgs inflation or the $R^2$ inflation in this level. 

\section{To the Effective $ R^2 $ Inflation} \label{sec:equiv}

\subsection{Relation between $ \xi $ and $ M $}

The degeneracy phenomenon implies that with a fixed $ \lambda=0.01 $, the amplitude of curvature perturbations, $ \mathcal{P}_{\mathcal{R}}(k)\simeq2\times10^{-9} $, gives a strong constraint on $ \xi $ and $ M $. 
\begin{figure}[h]
	\centering
	\includegraphics[width=.7\textwidth]{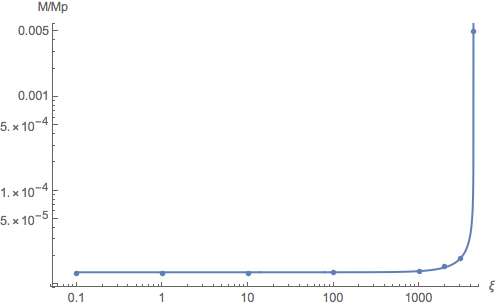}
	\caption{$ \lambda=0.01, \mathcal{P}_{\mathcal{R}}(k)\simeq2\times10^{-9} $~. The relation between $ \xi $ and $ M $ is given here. As can be seen, before $ \xi $ reaching $ 1000 $, $ M/M_p $ almost stays at around $ 1\times 10^{-5} $. However, when $ \xi >1000$, $ M/M_p $ begins to grow rapidly as $ \xi $ grows.}
	\label{fig:xi-m}
\end{figure}
Varying $ \xi $ from $ 0.1 $ to $ 4000 $, one obtains the relation depicted in Figure~\ref{fig:xi-m}. For small enough $ \xi $, $M$ remains almost constant at around $ 10^{-5}M_p $ which coincides with the case of the standard $R^2$ inflation. This situation holds until $ \xi $ reaches $ 1000 $ where $ M $ starts to grow rapidly to compensate the change of $ \xi $. In this logarithmic plot (Figure~\ref{fig:xi-m}), the relationship between $ M $ and $ \xi $ can be approximately separated into two branches. One is $ \xi\lesssim 1000 $ where we can just take the scalaron mass $ M $ to be the same as in the standard $R^2$ inflation. The other is $ \xi\gtrsim 1000 $ where the value of the scalaron mass in the $R^2$ inflation is no longer valid since the non-minimal coupling is so large that it modifies the model significantly. In order to maintain the amplitude of curvature perturbations, $ M $ must take a much larger value. 

So far we have qualitative understanding of this relation. We present more precise explanation in the following. 

\subsection{Explanation}
To explain this, we firstly have a look at the potential \eqref{potential}. In slow-roll regime, the Friedmann equation approximately gives 
\begin{align}\label{fe:smallxi}
	H^2\approx \frac{1}{3M^2_p} U(\psi, \chi)\approx \frac{1}{4} M^2 \left(1-e^{-\sqrt{\frac{2}{3}}\frac{\psi}{M_p}}\right)^2 
\end{align}
in large $ \psi $ regime with small $ \xi $ while 
\begin{align}\label{fe:largexi}
	H^2\approx \frac{1}{3M^2_p} U(\psi, \chi)\approx\frac{1}{4} M^2 \left(1-\frac{1}{M^2_p}\xi\chi^2e^{-\sqrt{\frac{2}{3}}\frac{\psi}{M_p}}\right)^2
\end{align}
in large $ \psi $ regime with large $ \xi $.  In the case of \eqref{fe:smallxi}, the second term in the parenthesis is negligible compared with unity so that Hubble parameter is just a constant completely determined by $ M $. However, in the case of \eqref{fe:largexi}, the second term in the parenthesis is not negligible compared with unity which means that the second factor in \eqref{fe:largexi} could possibly be much smaller than unity for large enough $ \xi $. In order to preserve the amplitude of curvature perturbations which is determined by Hubble parameter and its derivatives as mentioned above, we need a larger value of $ M $ to "protect" the Hubble parameter from being too small. 

Intuitively, the scalaron mass $ M $ mainly controls the height of the hill in the middle of the potential and while the non-minimal coupling $ \xi $ mainly controls the depth and position of valleys on both sides of the hill. For a given amplitude of curvature perturbations, we require the inflaton to slowly roll down a trajectory whose height is around a certain value during inflation. If $ M $ is not too large that means that the height of the central hill is small, the inflaton is allowed to roll along the central region of the potential, e.g. the left panel in Figure~\ref{fig:pot}. On the contrary, if $ M $ is too large, one then needs a larger value of $ \xi $ which would generate a deep valley on each side of the hill so that the inflaton can leave the too high hill top to roll along a trajectory which is of proper height to generate small enough curvature perturbations. 

One may see this relation more clearly without any conformal transformation by considering the relation between this two-field model and the $R^2$ inflation directly in the Jordan frame. From the action in this frame \eqref{action:j}, we find that the non-minimal coupling can be partially regarded as an extra contribution to the 
kinetic term for Higgs field $ \chi $ proportional to $ \xi $ since there are second derivatives in Ricci scalar so that we can realize this by using integration by parts. Then, for large $ \xi $ and $ \psi $ regime, the original kinetic term in \eqref{action:j} is negligible. Dropping the kinetic term for $ \chi $, the original action becomes 
\begin{align}
	S_{\text{J}}\approx\tilde{S_{\text{J}}}= \int d^4 x \sqrt{-\hat{g}} \left[\frac{M_p^2}{2} \hat{R}+\frac{1}{2}\xi \chi^2 \hat{R} +\frac{M_p^2}{12 M^2} \hat{R}^2 -\frac{\lambda}{4} \chi^4 \right]~.
\end{align}
As a result, the equation of motion for $ \chi $ just becomes a constraint on $ \chi $ \cite{Kehagias:2013mya} and $ R $ as $ \chi^2=\xi\hat{R}/\lambda $ which simplifies the action as 
\begin{align}
	\tilde{S_{\text{J}}} &=\int d^4 x \sqrt{-\hat{g}} \left[\frac{M_p^2}{2} \hat{R}+\left(\frac{M_p^2}{12 M^2}+\frac{\xi^2}{4\lambda}\right) \hat{R}^2 \right]\\
	&=\int d^4 x \sqrt{-\hat{g}} \left[\frac{M_p^2}{2} \hat{R}+\frac{M_p^2}{12 \tilde{M}^2}\hat{R}^2 \right]
\end{align}
where 
\begin{align}\label{meff:scalaron}
	\tilde{M}^2\equiv \frac{M^2}{1+3\xi^2 \frac{M^2}{\lambda M^2_p}}
\end{align}
is the effective mass squared of the scalaron, which should take $ \tilde{M}=1.3\times 10^{-5} M_p$ \cite{Starobinsky:1983zz,Faulkner:2006ub} to reproduce the observed amplitude of curvature perturbation. 

From \eqref{meff:scalaron} we can understand two characteristic regimes of Figure~\ref{fig:xi-m}. When the second term in the denominator is much smaller than unity, namely, $ \xi\ll \sqrt{\frac{\lambda}{3}}\frac{M_p}{M} \equiv \xi_c\cong 4.3\times 10^3 $, the scalaron mass should simply take the original value of $ R^2 $ model. As $ \xi $ increases, approaching the above critical value, $ M^2 $ also starts to increase according to 
\begin{align}
	M^2=\frac{\tilde{M}^2}{1-\frac{3\xi^2\tilde{M}^2}{\lambda M^2_p}}~,
\end{align}
reaching infinity at $ \xi=\xi_c $. This is in perfect agreement with Figure~\ref{fig:xi-m}. However, note that in the simplified case we do not have modification of sound speed. The effective scalaron mass can be regarded as the modification of potential that is different from the change of sound speed although they may both produce similar features on the power spectrum. 



As is well-known, only with either one of the two ingredients, $ R^2 $ or Higgs, it is enough to achieve successful inflation model with proper parameter value which is favored by the observation we have so far. Since the model we consider is the combination of these two single-field models, it goes back to either of them in some limits of the parameters. 

One can easily see from the Lagrangian that if we take $ \lambda\rightarrow 0 $ and $ \xi\rightarrow 0 $ in the model or just simply set $ \chi=0 $, what we get is just the $R^2$ inflation with only one parameter $ M $, the mass of the scalaron. The other limit is the Higgs inflation where we take $ M\rightarrow \infty $. Note that we cannot just take $ \psi=0 $ which is analogous to what we did above. The reason is that the new degree of freedom in $ \psi $ comes from the quadratic term of $ \hat{R} $ in \eqref{def:psi}. Only when $ \hat{R}^2 $ term vanishes, this "new" scalar field is completely determined by Higgs field, i.e. it is no longer a new degree of freedom. Thus, we go back to the Higgs inflation in this case with two parameters $ \xi $ and $ \lambda $.

\section{Conclusion and Outlook}\label{sec:concloutl}

In this paper, we have analyzed a two-field inflation model consisting of the $ R^2 $ term and the Higgs field in detail. This model can easily go back to the two single-field models, the $R^2$ inflation \cite{Starobinsky:1980te} and the Higgs 
inflation \cite{CervantesCota:1995tz,Bezrukov:2007ep,Barvinsky:2008ia}. 
We have considered the parameter space where $ \lambda=0.01 $ and $ \xi>0 $.  In the presence of mass hierarchy and considering slow-roll regime, one can integrate out the high energy part and obtain an effective single-field model with a slightly modified sound speed where we can easily calculate the power spectrum of curvature perturbations.  The modification of sound speed comes from the presence of turning in the inflation trajectory, but in our case it turned out to be negligibly small.  
For the amplitude of curvature perturbations to coincide with observation, we find that the predictions of this model are just the same as in the $R^2$ inflation or the Higgs inflation. Fixing the amplitude, $ \mathcal{P}_{\mathcal{R}}(k) $, we find a relation between the scalaron mass $ M $ and the non-minimal coupling $ \xi $ which helps us to notice the relation between this two-field model and the $R^2$ inflation directly in the Jordan frame. We can effectively regard this model in the parameter space considered as $ R^2 $ inflation with an effective scalaron mass which naturally explains the existence of a special relation between the two free parameters.

For typical values of the self coupling parameters of the standard Higgs field at high energy, this model 
gives essentially the same predictions as the $R^2$ inflation and the original Higgs inflation as far as the power spectrum is concerned.   These two models, however, have quite different reheating mechanisms
\cite{Starobinsky:1980te,St82,Mijic:1986iv,Gorbunov:2010bn,Arbuzova:2011fu}
with a much higher reheating temperature for the latter model with a possible violent behaviors due
to the non-minimal coupling \cite{Ema:2016dny}.   Since our model smoothly connects the
two limits, the number of $e$-folds, $N_\ast$, of
the pivot scale of CMB observation is also expected to shift from the value corresponding to
the pure Higgs model to that of $R^2$ model, which leads to an observational 
consequence \cite{Martin:2014nya}.  This shift, however, is degenerate to the expansion
history or the amount of entropy production after reheating which may be measured by the 
direct observation of high frequency tensor perturbations \cite{Nakayama:2008wy}.  

For smaller values of the self coupling than the case of the standard Higgs field with smaller
$\xi$, we may realize a situation both fields are in the slow roll regime and 
acquire non-negligible quantum fluctuations so that 
the isocurvature mode may also play an important role.   Furthermore, 
for $ \xi<0 $, we expect our model would have similar behaviors as in \cite{Pi:2017gih} where the two-field model can generate large fluctuations on small scale. 

Though we considered the model with only one scalar field, our results can be straightforwardly generelaized to an arbitrary number of mutually interacting scalar fields sufficiently strongly coupled to the Ricci scalar. 

\section*{Acknowledgements}

We thank the M. Sasaki, T. Suyama, Y. Watanabe, Y. Wang, Y. P. Wu, S. Pi, Y. Zhang, H. Lee for useful discussion. MH was supported by the Global Science Graduate Course (GSGC) program of the University of Tokyo. AS acknowledges RESCEU hospitality as a visiting professor. He was also partially supported by the RFBR grant 17-02-01008. The work of JY was supported by JSPS KAKENHI, Grant-in-Aid for Scientific Research 15H02082 and Grant-in-Aid for Scientific Research on Innovative Areas 15H05888.



\begin{thebibliography}{999}
	
	\bibitem{Starobinsky:1980te} 
	A.~A.~Starobinsky,
	Phys.\ Lett.\  {\bf 91B}, 99 (1980).
	doi:10.1016/0370-2693(80)90670-X
	
	\bibitem{Sato:1980yn} 
	K.~Sato,
	Mon.\ Not.\ Roy.\ Astron.\ Soc.\  {\bf 195}, 467 (1981).
	
	\bibitem{Guth:1980zm} 
	A.~H.~Guth,
	Phys.\ Rev.\ D {\bf 23}, 347 (1981).
	doi:10.1103/PhysRevD.23.347
	
	\bibitem{Linde:1981mu} 
	A.~D.~Linde,
	Phys.\ Lett.\  {\bf 108B}, 389 (1982).
	doi:10.1016/0370-2693(82)91219-9
	
	\bibitem{Albrecht:1982wi} 
	A.~Albrecht and P.~J.~Steinhardt,
	Phys.\ Rev.\ Lett.\  {\bf 48}, 1220 (1982).
	doi:10.1103/PhysRevLett.48.1220
	
\bibitem{Sato:2015dga} 
  For a review of single-field inflation, see e.g. K.~Sato and J.~Yokoyama,
  Int.\ J.\ Mod.\ Phys.\ D {\bf 24}, no. 11, 1530025 (2015).
  doi:10.1142/S0218271815300256
  
	
	
	\bibitem{Ade:2015lrj} 
	P.~A.~R.~Ade {\it et al.} [Planck Collaboration],
	Astron.\ Astrophys.\  {\bf 594}, A20 (2016)
	doi:10.1051/0004-6361/201525898
	[arXiv:1502.02114 [astro-ph.CO]].
	
\bibitem{CervantesCota:1995tz} 
  J.~L.~Cervantes-Cota and H.~Dehnen,
  Nucl.\ Phys.\ B {\bf 442}, 391 (1995)
  doi:10.1016/0550-3213(95)00128-X
  [astro-ph/9505069].
	
	\bibitem{Bezrukov:2007ep} 
	F.~L.~Bezrukov and M.~Shaposhnikov,
	Phys.\ Lett.\ B {\bf 659}, 703 (2008)
	doi:10.1016/j.physletb.2007.11.072
	[arXiv:0710.3755 [hep-th]].
	
\bibitem{Barvinsky:2008ia} 
  A.~O.~Barvinsky, A.~Y.~Kamenshchik and A.~A.~Starobinsky,
  JCAP {\bf 0811}, 021 (2008)
  doi:10.1088/1475-7516/2008/11/021
  [arXiv:0809.2104 [hep-ph]].
	
	\bibitem{Kamada:2012se} 
  K.~Kamada, T.~Kobayashi, T.~Takahashi, M.~Yamaguchi and J.~Yokoyama,
  Phys.\ Rev.\ D {\bf 86}, 023504 (2012)
  doi:10.1103/PhysRevD.86.023504
  [arXiv:1203.4059 [hep-ph]].
	
\bibitem{Kodama:1985bj} 
  H.~Kodama and M.~Sasaki,
  Prog.\ Theor.\ Phys.\ Suppl.\  {\bf 78}, 1 (1984).
  doi:10.1143/PTPS.78.1
	
	
	\bibitem{Chen:2009we} 
	X.~Chen and Y.~Wang,
	Phys.\ Rev.\ D {\bf 81}, 063511 (2010)
	doi:10.1103/PhysRevD.81.063511
	[arXiv:0909.0496 [astro-ph.CO]].
	
	\bibitem{Polarski:1994rz} 
	D.~Polarski and A.~A.~Starobinsky,
	Phys.\ Rev.\ D {\bf 50}, 6123 (1994)
	doi:10.1103/PhysRevD.50.6123
	[astro-ph/9404061].
	
	\bibitem{Ema:2016dny} 
	Y.~Ema, R.~Jinno, K.~Mukaida and K.~Nakayama,
	JCAP {\bf 1702}, no. 02, 045 (2017)
	doi:10.1088/1475-7516/2017/02/045
	[arXiv:1609.05209 [hep-ph]].
	
	\bibitem{Ema:2017rqn} 
	Y.~Ema,
	Phys.\ Lett.\ B {\bf 770}, 403 (2017)
	doi:10.1016/j.physletb.2017.04.060
	[arXiv:1701.07665 [hep-ph]].
	
	\bibitem{Wang:2017fuy} 
	Y.~C.~Wang and T.~Wang,
	Phys.\ Rev.\ D {\bf 96}, no. 12, 123506 (2017)
	doi:10.1103/PhysRevD.96.123506
	[arXiv:1701.06636 [gr-qc]].
	
	\bibitem{Maeda:1988ab} 
	K.~Maeda,
	Phys.\ Rev.\ D {\bf 39}, 3159 (1989).
	doi:10.1103/PhysRevD.39.3159
	
\bibitem{Maeda:1987xf} 
  K.~Maeda,
  Phys.\ Rev.\ D {\bf 37}, 858 (1988).
  doi:10.1103/PhysRevD.37.858
	
	
	\bibitem{Sasaki:1995aw} 
	M.~Sasaki and E.~D.~Stewart,
	Prog.\ Theor.\ Phys.\  {\bf 95}, 71 (1996)
	doi:10.1143/PTP.95.71
	[astro-ph/9507001].
	
	\bibitem{Achucarro:2010da} 
	A.~Achucarro, J.~O.~Gong, S.~Hardeman, G.~A.~Palma and S.~P.~Patil,
	JCAP {\bf 1101}, 030 (2011)
	doi:10.1088/1475-7516/2011/01/030
	[arXiv:1010.3693 [hep-ph]].
	
	\bibitem{Kehagias:2013mya} 
	A.~Kehagias, A.~Moradinezhad Dizgah and A.~Riotto,
	Phys.\ Rev.\ D {\bf 89}, no. 4, 043527 (2014)
	doi:10.1103/PhysRevD.89.043527
	[arXiv:1312.1155 [hep-th]].
	
	\bibitem{Pi:2017gih} 
	S.~Pi, Y.~l.~Zhang, Q.~G.~Huang and M.~Sasaki,
	arXiv:1712.09896 [astro-ph.CO].
	
	\bibitem{Achucarro:2012yr} 
	A.~Achucarro, V.~Atal, S.~Cespedes, J.~O.~Gong, G.~A.~Palma and S.~P.~Patil,
	Phys.\ Rev.\ D {\bf 86}, 121301 (2012)
	doi:10.1103/PhysRevD.86.121301
	[arXiv:1205.0710 [hep-th]].
	
\bibitem{Mijic:1986iv} 
  M.~B.~Mijic, M.~S.~Morris and W.~M.~Suen,
  Phys.\ Rev.\ D {\bf 34}, 2934 (1986).
  doi:10.1103/PhysRevD.34.2934
	
	\bibitem{Gorbunov:2010bn} 
  D.~S.~Gorbunov and A.~G.~Panin,
  Phys.\ Lett.\ B {\bf 700}, 157 (2011)
  doi:10.1016/j.physletb.2011.04.067
  [arXiv:1009.2448 [hep-ph]].
	
\bibitem{Arbuzova:2011fu} 
  E.~V.~Arbuzova, A.~D.~Dolgov and L.~Reverberi,
  JCAP {\bf 1202}, 049 (2012)
  doi:10.1088/1475-7516/2012/02/049
  [arXiv:1112.4995 [gr-qc]].
  
	 \bibitem{St82} A.~A.~Starobinsky, "Nonsingular model of the Universe with
	the quantum-gravitational de Sitter stage and its observational
	consequences", Proc. of the Second Seminar "Quantum Theory of gravity"
	(Moscow, 13-15 Oct. 1981), INR Press, Moscow, 1982, pp. 58-72; Quantum
	Gravity, eds. M~A.~Markov and P.~C.~West, Plenum Publ. Co., New York,
	1984, pp. 103-128.
	
	\bibitem{Martin:2014nya} 
	J.~Martin, C.~Ringeval and V.~Vennin,
	Phys.\ Rev.\ Lett.\  {\bf 114}, no. 8, 081303 (2015)
	doi:10.1103/PhysRevLett.114.081303
	[arXiv:1410.7958 [astro-ph.CO]].
	
	\bibitem{Starobinsky:1983zz} 
	A.~A.~Starobinsky,
	Sov.\ Astron.\ Lett.\  {\bf 9}, 302 (1983).
	
	\bibitem{Faulkner:2006ub} 
	T.~Faulkner, M.~Tegmark, E.~F.~Bunn and Y.~Mao,
	Phys.\ Rev.\ D {\bf 76}, 063505 (2007)
	doi:10.1103/PhysRevD.76.063505
	[astro-ph/0612569].
	
	\bibitem{Nakayama:2008wy} 
  K.~Nakayama, S.~Saito, Y.~Suwa and J.~Yokoyama,
  JCAP {\bf 0806}, 020 (2008)
  doi:10.1088/1475-7516/2008/06/020
  [arXiv:0804.1827 [astro-ph]].
  
  
\end{thebibliography}
\end{document}